\documentstyle[prl,aps]{revtex}        
\input epsf
\begin{document}
\draft
\twocolumn
\title{On the nature of the quantum states of macroscopic
systems} 
\author{Gyula Bene}
\address{Institute for Solid State Physics, E\"otv\"os University,
M\'uzeum krt. 6-8, H-1088 Budapest, Hungary}
\date{\today}
\maketitle


\begin{abstract}It is assumed that the quantum state that may describe
a macroscopic system at a given
instant of time is one of the eigenstates of the reduced
density matrix calculated from the wave function of the system plus its environment.
This implies that the above quantum state is a member of
a special
orthonormed set of states. Using a suitable
Monte-Carlo simulation, this property is shown to be
consistent with the extremely small standard deviation
for the coordinates and the momenta of macroscopic systems.  
Consequences for statistical mechanics and possible observable
effects are discussed.
\end{abstract}
\narrowtext

Despite of the well known conceptual difficulties\cite{cat},
 \cite{EPR}, \cite{Bell}, there is no compelling reason to believe
that quantum mechanics fails when applied to macroscopic systems. It is then
reasonable to ask what are the properties of quantum states of macroscopic
systems. In order to fit experience, such a quantum state should yield extremely small
standard deviation for both the position and the momentum of the center of mass.
On the other hand, due to the large number of particles contained in macroscopic bodies,
they possess an extremely high energy level density, or an extremely small average level spacing.
Correspondingly, a macroscopic system $M$ cannot be isolated from its environment 
$E$\footnote{The environment
itself is a physical system, consisting of air molecules, radiation field, other macroscopic systems
etc.}, hence, according to orthodox quantum mechanics, it cannot be characterized by
a wave function, only by a suitable reduced density matrix $\rho$, resulting from the
wave function $\Psi$ of the composite system $M+E$ 
by taking the partial trace of $|\Psi><\Psi|$ over the Hilbert space of the environment.\cite{Wigner}

Nevertheless, it is still possible to assign a wave function to the system $M$.
To show this, let us consider a simplified description of a spin measurement on a spin-half
particle.
Suppose that the $S_z$ spin component is measured, and initially the particle
is in the state $|\uparrow>$, which is the eigenstate of the $\hat S_z$
operator corresponding to the eigenvalue $+\frac{\hbar}{2}$. Then the measurement
process may be symbolically written as
$
|\uparrow>|m_0>\, \rightarrow \, |\uparrow>|m_\uparrow>$,
where $|m_0>$ stands for the initial state of the measuring device, 
and $|m_\uparrow>$ stands for the state of the measuring device
 when it shows the result $+\frac{\hbar}{2}$. The horizontal arrow denotes the (unitary) 
 time evolution, satisfying the Schr\"odinger equation. 
 
 Similarly, if the initial state of the particle is $|\downarrow>$, 
 corresponding to the eigenvalue $-\frac{\hbar}{2}$, the measurement process
 can be written as
$
|\downarrow>|m_0> \,\rightarrow \, |\downarrow>|m_\downarrow>$.

If the initial state of the particle is the superposition 
$\alpha |\uparrow> +\beta |\downarrow>$ (with $|\alpha|^2+|\beta|^2=1$),
then, owing to the linearity of the Schr\"odinger equation we may write
$
\left(\alpha |\uparrow> +\beta |\downarrow>\right) |m_0> \rightarrow 
\alpha |\uparrow>|m_\uparrow> +\beta |\downarrow>|m_\downarrow>$.

Looking at the final state of the system, one can see that the measuring device 
has no own wave function, instead, it can be characterized by the 
density matrix 
$
\hat \rho=|m_\uparrow>|\alpha|^2 <m_\uparrow|
+|m_\downarrow>|\beta|^2 <m_\downarrow|$.

Certainly, in such a situation the result of the measurement is either
$+\frac{\hbar}{2}$, corresponding to the state $ |m_\uparrow>$, 
or $-\frac{\hbar}{2}$, corresponding to the state $|m_\downarrow>$. 
As the states $ |m_\uparrow>$ and $|m_\downarrow>$ are orthogonal, they
are just the eigenstates of $\hat \rho$. Extrapolating the conclusion of this
simple example, it is reasonable to assume that a macroscopic system is described by one of the
eigenstates of its reduced density matrix $\hat \rho$, calculated from the wave function
of the Universe, and the probability that this particular eigenstate occurs
is given by the corresponding eigenvalue. 
This has already been proposed by several authors\cite{Zeh}, \cite{Albrecht},
\cite{Dieks}. A consistent theory involving this idea can be found 
in Ref.\cite{Bene}. It is of worth mentioning that this latter theory is
an explicit counterexample to Bell's theorem\cite{Bell}.

The next question is how the reduced density matrix of a macroscopic system 
looks like if the interaction with the environment is taken into account.
This is the central problem of the phenomenon 'environment induced
decoherence'\cite{Zeh}, \cite{Zurek}, and as such, has been much studied in model systems.
A typical model\cite{Caldeira-Leggett} contains a number of harmonic oscillators playing the role
of a heat bath (this is the environment) and a macroscopic object coupled to it.
As a result of the interaction, the reduced density matrix of the macroscopic object 
becomes nearly diagonal
in both coordinate and momentum representation, so that the width across the diagonal
is microscopically small, while  the distribution along the diagonal (this is the
probability distribution of the coordinate in coordinate representation and 
the probability distribution of the momentum in momentum representation) extends to a
macroscopic regime. Note that (up to the present author's best knowledge)
none of the decoherence models studied so far possesses the property that 
the eigenstates of the reduced density matrix are generically localized. 
The reason may well be that these models are oversimplified.

Considering now the mathematical consequences of the above idea, one finds
that the quantum state of a macroscopic system is a member of an
orthonormed set of states\footnote{The latter property is due to
the hermiticity of the reduced density matrix $\hat \rho$.}. 
The relevant eigenstates of the reduced density matrix
(i.e.,  those which occur with non-negligible probability) 
must correspond
to a macroscopic region of the classical phase space, therefore,
their number is given by this phase space volume divided by Planck's
constant $h$ (in one dimension, in $f$ dimension one must use
$h^f$ instead).
All the relevant eigenstates must possess the classical property 
that the standard deviation
for both the coordinates and the momenta ($\delta x$ and $\delta p$, respectively) 
is microscopically small. The aim of the
present paper is to answer the question whether it is possible mathematically 
at all. 

It is natural to ask whether there exists such a complete, orthonormed basis 
that the standard deviations $\delta x$, $\delta p$ for {\em all} the basis functions
lie below a common bound. The answer is known in the one-dimensional 
case when the basis functions
are generated from the same function by translations along the $x$ and $p$
directions\cite{Balian}. It turns out that completeness, orthonormality 
and boundedness of $\delta x$ and $\delta p$ cannot be simultaneously fulfilled. 

We made a numerical study to decide whether it is also true if the basis 
functions are not generated by translations. The method was the following.
Initially one dimensional oscillator eigenstates $|n>$ below a given energy 
$E_0=(N-1/2)\hbar \omega$ were considered. Certainly, in this case 
$\delta x_n \propto \delta p_n \propto \sqrt{n}$. Then new basis functions 
$|\tilde n>$ were
constructed by a unitary transformation
$
|\tilde n> =\sum_{j=0}^{N-1} U_{n j} |j>$.
 
The transformation matrix was chosen so as to minimize 
the sum of $\delta x^2 + \delta p^2$ for all the $N$ basis functions.
(The units were chosen so that
$\hbar=1$, $\omega=1$, $m=1$.) 
As 
$
\delta x^2=<\tilde n|\hat x^2|\tilde n>-<\tilde n|\hat x|\tilde n>^2$,
and
$\sum_{n=0}^{N-1} <\tilde n|\hat x^2|\tilde n>$\hfill\break \mbox{$=
\sum_{n=0}^{N-1}\sum_{j=0}^{N-1}\sum_{k=0}^{N-1} U^*_{n,j} 
U_{n,k}<j|\hat x^2|k>$} \mbox{$=\sum_{j=0}^{N-1} <j|\hat x^2|j>$}
(similarly for $\hat p$), it was enough to maximize the quantity
$
S=\sum_{n=0}^{N-1} <\tilde n|\hat x|\tilde n>^2$ \mbox{$+<\tilde n|\hat p|\tilde n>^2$}.

The transformation matrix was built up as a product
of unitary matrices, each having only one nontrivial 2 by 2 block. The
corresponding indices were randomly chosen. For the other indices
the matrix was just the unit matrix. The most general 2 by 2 unitary
transformation is the transformation matrix of the basic spinors
which may be parametrized by three angles. For each matrix, these angles
were chosen at random, and it was subsequently 
considered whether the result of this
transformation gave rise to a larger sum $S$ or not. If yes, then
the transformation was kept, if not, it was discarded. The method
was therefore a simple Monte-Carlo simulation. The procedure was
continued until the quantity $S$ saturated. The resulting average deviations 
(i.e., the sum of $\delta x^2 + \delta p^2$ for all the $N$ states 
divided by $N$) as a function of various $N$ values 
is plotted in Fig.1. The error of the data is less than $10^{-2}$.
 The measured values are surprisingly well fitted
by the logarithmic function $1.0+0.6\,\ln\left(N\right)$. This result
supports the generalization of the statement of Ref.\cite{Balian}, i.e., 
it shows that no complete, orthonormed basis with bounded
standard deviations $\delta x$, $\delta p$ exists. 
\begin{figure}
\hspace*{-1.5cm}
\epsfbox{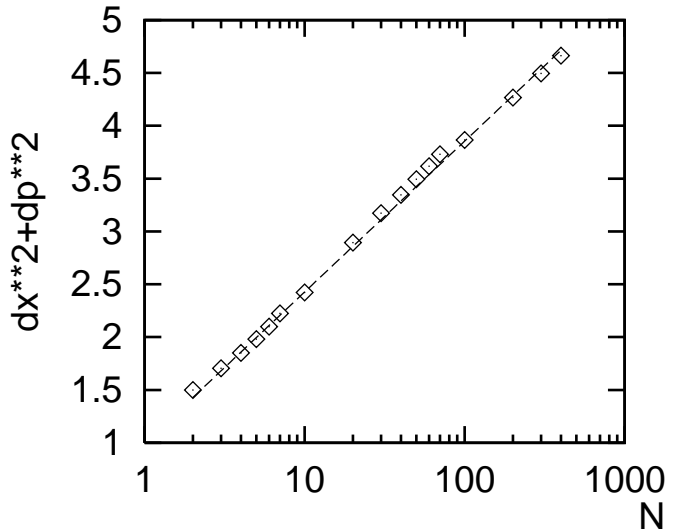}
\caption{Diamonds: the average of $\delta x^2 + \delta p^2$ over all the $N$ states versus $N$. Dashed line:
the logarithmic fit $1.0+0.6 \log(N)$.}
\end{figure}

Returning to the original question, we show that the logarithmic
dependence on $N$ is consistent with the requirement that 
the relevant eigenstates (i.e., those with significant probability to occur) 
of the reduced density matrix 
must have small $\delta x$, $\delta p$ values\footnote{The irrelevant eigenstates may be
identified with the original oscillator eigenstates 
$|n>$ with $n\ge N$.}. Indeed, the relevant 
eigenstates of $\hat \rho$ are associated with a large but finite
portion of the classical phase space. We choose the number $N$ to be
the ratio of this phase space volume and Planck's constant $h$.\footnote{This 
is valid in one dimension. In dimension $d$ we divide by $h^d$, and 
the number $N$ is chosen to be the $d$-th root of the ratio, as the
basis functions will be constructed as a $d$-fold direct product 
of the one-dimensional basis. For simplicity we shall
consider only the one-dimensional case. }
 The widths $\delta x$, $\delta p$ of the relevant
eigenstates are proportional to $\sqrt{0.5+0.3 \ln\left(N\right)}$. This quantity
remains moderate even for an astronomically large $N$ value. As an example,
consider the motion of the Earth. The linear size in coordinate space is
$3\times 10^{11}$ m, and that in momentum space is $3.6\times 10^{29} \rm kg\frac{m}{s}$.
Therefore, $N=1.6\times 10^{74}$ and $\sqrt{0.5+0.3\ln\left(N\right)}=7.2$. 
Thus the widths $\delta x$, $\delta p$ are only one order of magnitude
larger than those of the minimal wave packet $|0>$. As we see, the
logarithmic dependence makes the above assumption about the wave functions
of macroscopic systems physically acceptable. 
Nevertheless, the problem is left open
whether these states actually occur in Nature 
and if so, what the origin of their 'localization' property is.
Certainly, it should be due to the interaction between the macroscopic system
and its environment (as this produces the mixed state described by $\hat \rho$),
but that feature of the dynamics
which is responsible for classical behavior is unknown. It should be
emphasized that the 'nearly diagonal' shape of $\hat \rho$ in coordinate
and in momentum representation is insufficient. E.g., if $\rho(x,x')\propto
\exp\left(-\frac{(x-x')^2}{2\sigma_1^2}-\frac{(x+x')^2}{2\sigma_2^2}\right)$
with $\sigma_1<<\sigma_2$, the density matrix is nearly diagonal in
both coordinate and momentum representation, but its eigenstates
are oscillator eigenstates with $\delta x_n\propto \delta p_n \propto \sqrt{n}$.

Here we consider only some consequences if the above 'logarithmic
localization' of the macroscopic quantum states actually occurs.

Let us consider first a harmonic oscillator of unit mass and frequency
(certainly, these units may be as large as 1 kg and 1 s$^{-1}$),
 and calculate the
mean deviation of the energy in the above localized states.
Certainly, the average of this quantity depends on the number of the states
($N$), as for larger $N$ the states are broader. The dependence of the
average square deviation of the energy on the number of states is
plotted in Fig.2. 
\begin{figure}
\hspace*{-1.5cm}
\epsfbox{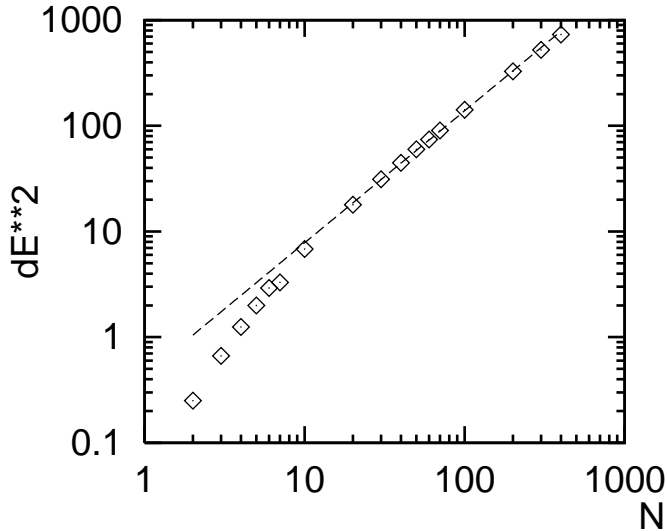}
\caption{Diamonds: the average of $\delta E^2$ (the squared deviation of the energy) over all the $N$ states versus $N$. Dashed line:
the fit $0.44\,N^{1.25}$}
\end{figure}
Somewhat surprisingly, this dependence is not 
logarithmic, but power-like, with the exponent $\approx 1.25$. This means
that
\begin{eqnarray}
\delta E \propto N^{0.63}\label{exx}
\end{eqnarray}
The mean energy is proportional to $N$ (as $N$ is chosen 
according to the relevant classical phase space volume),
therefore, the relative uncertainty of the energy, i.e., 
$\delta E/E$ goes to zero as 
$N^{-0.39}\propto \left(\frac{E}{\hbar \omega}\right)^{-0.39}$. 
Certainly, in case of a macroscopic oscillator this is
unobservable, although $\delta E$ itself can be quite large
at the microscopic scale $\hbar \omega$. 

In order to understand the behavior (\ref{exx}), let us consider
the asymptotic behavior of the above localized states in coordinate
representation. Beyond $\sqrt{N}$ they fall off as
$\exp\left(-x^2/2\right)$, due to their construction. For smaller distances (down to some
$x_0$)
they decay like a power. The exponent is $\nu\approx -1.5$.
This readily implies that 
\begin{eqnarray}
\delta^2 x=\int_{-\infty}^\infty dx |\psi(x)|^2 x^2
\approx {\rm const.}\times \int_{x_0}^{\sqrt{N}} dx\; x^{-1}\nonumber\\
={\rm a_1} + {\rm b_1} \ln\left(N\right)\nonumber
\end{eqnarray}
where ${\rm a_1}$ and ${\rm b_1}$ are constants.
The behavior in momentum space is the same.
As for the square deviation of the energy, we have to consider
the expectation value of $\hat x^4+\hat p^4+\hat x^2\hat p^2 
+\hat p^2\hat x^2$. We may write e.g.
\begin{eqnarray}
<\hat x^4>=\int_{-\infty}^\infty dx |\psi(x)|^2 x^4
\approx {\rm const.}\times \int_{x_0}^{\sqrt{N}} dx\; x\nonumber\\
={\rm a_2} + {\rm b_2} N\nonumber
\end{eqnarray}
One can derive the same linear estimate for $<\hat p^4>$ as well.
This implies that $\delta^2 E$ grows at least linearly with $N$, 
which is consistent with the numerically obtained $N^{1.25}$.

In order to get further consequences, 
let us consider a macroscopic system being in thermal
equilibrium with a heat bath at temperature $T$. What is 
the density matrix of that system at a given instant of time?
One would guess that it is  
\begin{eqnarray}
\frac{1}{Z}\exp\left(-\beta \hat H\right)\,,\label{e9}
\end{eqnarray}
where $\beta=\frac{1}{k_B T}$ and $Z$ stands for the canonical
partition function. This expression, however, substitutes
a long time average, therefore, one cannot expect that it is
strictly valid at a given instant of time. According to the previous considerations,
it is more realistic to assume that (\ref{e9}) gives only the long time average
of the density matrix, which is actually given by
$
\hat \rho = \sum_n |\varphi_n>p_n <\varphi_n|\,,\label{e10}
$
where $|\varphi_n>$ is narrow in both coordinate and in
momentum space if $p_n$ is not negligibly small.
As energy eigenstates are generically not localized,
the $|\varphi_n>$-s are not eigenstates of the Hamiltonian $\hat H$,
therefore, $\hat \rho$ will have nondiagonal matrix elements
in energy representation. Explicitly, these matrix elements
are 
$
\rho_{j k}=<\xi_j|\hat \rho|\xi_k>
=\sum_n p_n <\xi_j|\varphi_n><\varphi_n|\xi_k>$,
where $|\xi_k>$-s stand for the energy eigenstates.
The states $|\varphi_n>$ are also localized in energy representation
(i.e., for a given $n$ $<\xi_j|\varphi_n>$ takes on significant
values only for a narrow range of the energy eigenvalues 
$E_j$), thus $\rho_{j,k}$ is a narrow band matrix.

The existence of nondiagonal matrix elements in energy representation
can in principle lead to experimentally observable effects. 
Suppose one isolates the macroscopic system in question so that
the interaction energy $\Delta E_I$ 
with the environment becomes much less than
the effective width $\delta E$ of the density matrix in energy representation.
Under such circumstances consider the linear response of the system. 
For a time $t$ which is much less than 
$\hbar/\Delta E_I$ 
but much larger than
$\hbar/\delta E$ 
the deviation in the expectation value of an operator 
$\hat A$ is 
$$\delta<\hat A>(t)=-\frac{i}{\hbar}\int_0^t dt' 
{\rm Tr}\left(\left[\hat A(t),\hat H_1(t')\right]
\hat \rho(0)\right)\,,$$
where $\hat H_1$ is the perturbation, and the operators are Heisenberg operators
calculated with the unperturbed Hamiltonian $\hat H_0$. Choosing $\hat A=\hat H_0$,
the response can be cast to the form 
\begin{eqnarray}
\delta<\hat H_0>(t)=\sum_{j,k} \left(\rho_{j,k}(t)-\rho_{j,k}(0)\right)
<\xi_k|\hat H_1(0)|\xi_j>\,,\nonumber
\end{eqnarray}
where $\rho_{j,k}(t)=\exp\left(-\frac{i}{\hbar}(E_k-E_j)t\right)\rho_{j,k}(0)$.
If the density matrix were strictly diagonal in energy representation, the response
$\delta<\hat H_0>(t)$ would vanish. Actually, as noted above, 
$\rho_{j,k}(0)$ is a narrow band matrix, hence
the Fourier transform of the response vanishes at high frequences but
becomes significant at low frequences. Therefore, 
the localization property of the eigenstates 
shows up in a suitable linear response which increases at low frequencies. 
It is a challenging question whether this behavior may be related to
the celebrated $1/f$ noise\cite{1pf}. It is also of worth noting
that the deviation of the density matrix from the canonical expression
(\ref{e9}) implies the breakdown of the fluctuation-dissipation theorem.

In conclusion, we have seen that \hfill\break
i, in one dimension there is no localized basis in the strict sense \hfill\break
ii, the numerically found 'logarithmically localized' basis 
demonstrates that the idea that macroscopic systems are
described by the eigenstates of their reduced density matrix
is consistent with experience\hfill\break
iii, this idea implies that the density matrix of a macroscopic
system being in thermal equilibrium is not diagonal in energy 
representation, that may be observed
in the spectrum of a suitable linear response function
as an enhancement at low frequences.

\section{Acknowledgements}
The author is obliged to Prof.A.Voros for an enlightening discussion
and also for reference \cite{Balian}.

This work has been partially supported by the Hungarian Aca\-demy of 
 Sciences
 under Grant Nos. OTKA T 017493, OTKA F 17166 and OTKA F 019266, 
 by the Hungarian Ministry of Culture and Education 
 under Grant No. FKFP 0159/1997 and
 by the German-Hungarian Scientific and Technological Cooperation Project No. D 125.
 
 The author wants to thank for the hospitality of the {\em Institut f\"ur 
 Festk\"orperphysik, Forschungszentrum J\"ulich GmbH} and of the
 {\em Lehrstuhl f\"ur Theoretische Physik, Universit\"at zu K\"oln} (at the latter department
 the author stayed as a DAAD fellow) where 
 important parts of the work have been done.

\end{document}